\documentclass[12pt]{article}

\usepackage[body={17cm, 23cm, centered}]{geometry}
\usepackage{epsfig,verbatim,cite}
\usepackage{lmodern}
\usepackage{braket}
\usepackage[T1]{fontenc}
\usepackage[utf8]{inputenc}			

\usepackage{mathtext,marvosym,textcomp}
\usepackage{mathtools}
\usepackage{slashed}
\usepackage[dvipsnames,svgnames,table]{xcolor}
\usepackage{tikz}
\usepackage[ normalem]{ulem}
\usepackage{tocloft}
\usepackage{epsfig,amsmath,amsfonts,amssymb,arydshln}
\usepackage[makeroom]{cancel}
\usepackage{multirow}

 \usepackage[debug,pageanchor=false]{hyperref}
\hypersetup{colorlinks=true,linktocpage,breaklinks,
            urlcolor=blue,
            linkcolor=teal,
            citecolor=purple
            }

\numberwithin{equation}{section}

\def\a{\alpha} 
\def\b{\beta} 
\def\g{\gamma} 
\def\d{\delta} 
\def\e{\epsilon}
 
\def\z{\zeta} 
\def\h{\eta}

\def\l{\lambda} 
\def\m{\mu}
\def\n{\nu} 
\def\x{\xi} 
\def\p{\pi}
\def\r{\rho}
\def\q{\theta}
\def\s{\sigma} 
\def\t{\tau}  
\def\f{\phi}

\def\W{\Omega}

\def\fr{\frac}

\def\dt{\partial}

\def\ph{\phantom}

\def\mc{\mathcal}

\def\mH{\mathcal{H}}

\def\mH{\mathcal{H}}

\def\tX{\tilde{X}}

\def\SS{\mathbb{S}}
\def\TT{\mathbb{T}}

\def\AdS{\mathrm{AdS}}

\def\rmSO{\mathrm{SO}}

\def\rmO{\mathrm{O}}

\def\frso{\mathfrak{so}}

\def\diag{\mathrm{diag}}

\def\Tr{\mathrm{Tr}}
\def\Pf{\mathrm{Pf}}

\usepackage[pdftex]{pict2e}
\usepackage[dvipsnames]{xcolor}

\usepackage{tcolorbox}

\begin{document}

\begin{titlepage}
\ph{preprint}

\vfill

\begin{center}
   \baselineskip=16pt
   {\large \bf Brane-probes response to poly-vector deformations
   }
   \vskip 1cm
    Sergei Barakin$^{a,c}$\footnote{\tt barakin.serge@gmail.com },
    Angelina Kurenkova$^a$\footnote{\tt angelinakurenkova16108@gmail.com},
    Edvard T. Musaev$^{b,c,a}$\footnote{\tt emusaev@theor.jinr.ru}
       \vskip .6cm
             \begin{small}
                          {\it
                          $^a$Institute of Theoretical and Mathematical Physics, Moscow State University, 119991, Russia
                          $^b$Bogoliubov Laboratory of Theoretical Physics, Joint Institute for Nuclear Research, \\ 6, Joliot Curie, 141980 Dubna, Russia\\
                          $^c$Moscow Institute of Physics and Technology, 
                          Laboratory of High Energy Physics, \\
                          9, Institutskii pereulok, 141702, Dolgoprudny, Russia
                          } \\ 
\end{small}
\end{center}

\vfill 
\begin{center} 
\textbf{Abstract}
\end{center} 
\begin{quote}
We probe poly-vector deformations of Type II backgrounds using the fundamental string, D0-brane and D3-brane, and of the 11D membrane background by the fundamental M2-brane. We show that the corresponding deformations of the world-volume theories can be written in terms of flows similar to the $\mathrm{T}\bar{\mathrm{T}}$ flow (for the D3-brane flow --- with a square-root term). This interpretation works equally well for abelian (TsT) and non-abelian deformations.
\end{quote} 

\vfill
\setcounter{footnote}{0}
\end{titlepage}

\tableofcontents

\setcounter{page}{2}

\section{Introduction}

Due to background dependence the properties of string theory may change drastically when moving along the space of supergravity solutions. One famous example is provided by backgrounds with critical electric fields, where the string spectrum becomes non-relativistic and one is naturally led to the NRST/NCOS regime \cite{Seiberg:2000ms,Seiberg:2000gc}.  As it has been discussed in \cite{Barakin:2025jwp,Barakin:2026mxz} poly-vector deformations provide a unified formalism to parametrize such a movement both in 10 and 11 dimensions. When speaking of string dynamics a natural question arises of how the probe string reacts to such deformations of the underlying geometry. For TsT (equivalently, abelian bi-vector) deformations the answer is known and is remarkably fascinating: the $\mathrm{T}\bar{\mathrm{T}}$ deformations of the string sigma-model may be viewed as the world-sheet manifestation of a target-space bi-vector transform \cite{Giveon:2017nie,Asrat:2017tzd,Chakraborty:2018vja,Araujo:2018rho}. Moreover, in \cite{Blair:2020ops} it was shown that such deformations naturally interpolate between relativistic and non-relativistic descriptions. This observation strongly suggests that the deformation parameter should be understood not only as coupling-space data of the world-sheet theory, but also as a coordinate on a larger moduli space of backgrounds.

In 2D CFT $\mathrm{T}\bar{\mathrm{T}}$ deformation, as an example of irrelevant deformations generated by composite operators, was first revealed by Zamolodchikov's factorization formula and then developed into a finite-coupling flow in \cite{Zamolodchikov:2004ce,Smirnov:2016lqw}. The remarkable feature of this flow is that despite being irrelevant in the renormalization-group sense, it retains a huge degree of analytic control: the finite-volume spectrum satisfies a Burgers-type equation, integrability is often preserved, and the deformation admits a geometric interpretation in terms of gravitational dressing and finite-cut-off holography (for double-trace deformations) and changing the AdS geometry (for single-trace deformations) \cite{Cavaglia:2016oda,Dubovsky:2018bmo,McGough:2016lol,Araujo:2018rho,Guica:2019nzm,Ferko:2025elh}. These features have made $\mathrm{T}\bar{\mathrm{T}}$ a natural framework for studying UV-complete irrelevant flows, effects of non-locality compatible with solvability, and the relations between integrable QFT, string theory, and holography.

One becomes interested in extending this picture in two directions: beyond two dimensions and beyond abelian bi-vectors. In higher dimensions the special factorization behind the original $\mathrm{T}\bar{\mathrm{T}}$ construction is no longer available in the same universal form, and therefore one should not expect a unique generalization. Nevertheless, a number of higher-dimensional analogues have been proposed, often motivated by determinant-type operators, holography, or brane effective actions \cite{Cardy:2018sdv,Taylor:2018xcy,Hartman:2018tkw}. In this broader picture exact statements are known in special sectors: at large $N$, in holographic settings, and for particular world-volume systems.  Of particular relevance for the present work are the analyses of \cite{Blair:2020ops,Blair:2024aqz}, where higher-dimensional analogues of $\mathrm{T}\bar{\mathrm{T}}$-like flows were related to M2-brane dynamics and to the interpolation between relativistic and non-relativistic regimes.  These results suggest that in dimensions greater than two one should perhaps think not only in terms of composite stress-tensor operators, but also in terms of deformations of the target-space background seen by the corresponding probe.

A framework in which the idea of generalization of bi-vector deformations becomes systematic is the poly-vector deformations formalism suggested in \cite{Bakhmatov:2019dow} and developed in \cite{Gubarev:2020ydf}. In that approach, deformations are encoded by poly-vectors of various ranks, placing bi-vector deformations and their U-duality analogues on the same footing. Analysis of the recent works \cite{Barakin:2025jwp,Barakin:2026mxz} shows that certain poly-vector deformations can be interpreted as motions in string/M-theory moduli space between different kinematical regimes. In particular, bi-vector deformations add dissolved F1 charge, tri-vector deformations add dissolved M2 charge, quadri-vector deformations add D3-brane charge, and uni-vector deformations in the Type IIA setup are naturally tied to D0 charge and, in the critical limit, to DLCQ. From this point of view poly-vector deformations interpolate between regimes that are naturally interpreted as relativistic, non-relativistic, or light-cone. This makes it natural to ask a complementary question: what dynamics do the fundamental probes themselves exhibit when placed on such deformed backgrounds? 

In this paper we address this question by considering the fundamental string, the D0- and D3-brane and the M2 membrane on deformations of their own backgrounds. More concretely, we place the fundamental string on a bi-vector deformed string background, the D0-brane on a uni-vector deformed D0-brane background, the D3-brane on a quadri-vector deformed D3-brane background, and the M2-brane on a tri-vector deformed M2 background. In all these cases the rank of the poly-vector matches the dimensionality of the probe world-volume, and one is naturally led to flow equations for the corresponding deformed solutions. In the D0 and the membrane case we find agreement with the structures discussed in \cite{Blair:2024aqz,Blair:2020ops}, although in our setup the flow carries an additional factor of the harmonic function $H(r)$. This extra factor has a simple origin in the fact that we work directly with the probe on the full curved background. We consider both abelian and non-abelian deformations, and find that in the latter case one may still write a proper flow equation with a different composite operator.

The rest of the paper is organised as follows. In Section~\ref{sec:II} we derive the flow equation for the fundamental string on the bi-vector deformed F1  and $\AdS_3\times \SS^3 \times TT^4$ backgrounds; of the D3-brane on the D3-brane deformed background and of the D0-brane on the D0-brane deformed background. In Section~\ref{sec:M2} we derive the membrane flow and compare it to \cite{Blair:2020ops,Blair:2024aqz}. In Section~\ref{sec:coord} we discuss the interpretation of bi-vector and uni-vector deformations as coordinate transformations in a parent theory, observed earlier in \cite{Gubarev:2025hvr}, in relation to flow equations. We conclude in  Section~\ref{sec:conclusions} with a discussion of the relation between $\mathrm{T}\bar{\mathrm{T}}$-type deformations, poly-vector deformations, coordinate transformations and exact solvability.

\section{Type II probes}
\label{sec:II}

In this section we consider the fundamental string and the D3-brane probing Type II backgrounds deformed by bi-vectors and quadri-vectors respectively. For the latter the formalism was developed in \cite{Gubarev:2024tks} (for backgrounds of a particular form suitable for our purposes). In \cite{Blair:2020ops} it was shown that $\mathrm{T}\bar{\mathrm{T}}$ deformations of the Nambu--Goto string on flat space-time produce the non-relativistic string (in a certain limit). On the other hand, in \cite{Barakin:2025jwp} it has been shown that starting from the F1 background and performing abelian deformations with $\b \sim \dt_0\wedge \dt_1$ along the world-volume directions one effectively changes the core charge of the background. Applied to a Dp-brane background such a deformation induces a dissolved string charge on the D3-brane world-volume producing the background of the Dp-F1 bound state, that stands for a natural description of Dp-brane backgrounds in NRST \cite{Harmark:2000wv}.

We are now interested in analysis of the fundamental string dynamics on the F1 background deformed by abelian and non-abelian bi-vectors. As we will see, both cases can be packed into a nice form of a flow triggered by a composite operator, that depends on the deformation chosen.

\subsection{Deformations of the string background}

Start with the fundamental string background, that is a solution to Type II supergravity equations of the form
\begin{equation}
\label{eq:F1m}
\begin{aligned}
    ds^2 & = H(r)^{-1} \big(- dx_0^2 + dx_1^2 \big) + dr^2 + r^2 ds^2_{\SS^7} \\ 
    B_2 & = H(r)^{-1} dx^0\wedge dx^1,\\
    e^{-2\f} & = e^{-2\f_0}H(r), \\
    H(r) & = 1 + \fr{R^6}{r^6}
\end{aligned}
\end{equation}
Near horizon the background looks like a warped $\AdS_3$ times a 7-sphere 
\begin{equation}
\label{eq:F1_NH}
    \begin{aligned}
        dds^2 & = \fr{1}{z^3}\big(-dx_0^2+dx_1^2\big) + \fr{R^2 dz^2}{4 z^3} + \fr{R^2}{z}d\W_7^2, \\
        B_2 & = \fr{1}{z^3}dx^0 \wedge x^1\\
        e^{-2\f} & = e^{-2\f_0}z^3,
    \end{aligned}
\end{equation}
where we introduce the radial coordinate in $\AdS_3$ as $z= R^2/r^2$.

To perform bi-vector deformations we use the standard formula
\begin{equation}
    (g+b)^{-1} = (G+B)^{-1} + \b,
\end{equation}
where $G$, $B$ correspond to an undeformed solution. In our case, however, the matrix $G+B$ is degenerate in the block $\{01\}$. To override that we consider a gauge transformed Kalb--Ramond field $B_2 \to B_2 + \e dx^0 \wedge dx^1$ with constant $\e$, that is sent to zero after the deformation. Deformations will be performed along the world-volume Killing vectors of the fundamental string background that are the momenta and one boost:
\begin{equation}
    P_a, \quad M = x^0 \dt_1 + x^1 \dt_0=\e^{ab}\s_a \dt_b. 
\end{equation}
It appears useful to include also the vector
\begin{equation}
    D = x^0 \dt_0 + x^1 \dt_1 + z \dt_z,
\end{equation}
that is a conformal Killing vector of the full solution and the standard Killing vector of $\AdS_3$.

Consider now deformations constructed from these vectors sorting by powers of coordinates in $\beta$.

\textbullet\;PP-deformation is an abelian deformation given by the bi-vector
\begin{equation}
    \b = \fr12\g \dt_0\wedge \dt_1.
\end{equation}
This deformation was considered in \cite{Barakin:2025jwp}, where it was shown that it leaves the form of the string background intact replacing the harmonic function as
\begin{equation}
    H_\g(r) = H(r) + \g.
\end{equation}
Important for construction of the corresponding flow is that $\dt_\g H_\g(r) = $ const. Applied to the near-horizon limit of the F1 solution one gets the full string background.

\textbullet\;PM-deformation is non-abelian and the corresponding bi-vector in general reads
\begin{equation}
    \b = \fr12 \x^a P_a \wedge M = \fr12\x_a x^a \dt_0 \wedge \dt_1.
\end{equation}
This deformation shifts the harmonic function as $H(r) \to H(r) + \x_a x^a$, and applied to the near-horizon limit on gets the following solution
\begin{equation}
    \begin{aligned}
        ds^2 & = \fr{\h_{ab}dx^a dx^b}{z^3 - \x_a x^a} + \fr{R^2 dz^2}{4 z^3} + \fr{R^2}{z}d\W_7^2, \\
        B_2 & = \fr{dx^0 \wedge dx^1}{z^3 - \x_a x^a}\\ 
        e^{-2\f}& = z^3 - \x_a x^a
    \end{aligned}
\end{equation}

It is worth mentioning, that the classical Yang--Baxter equation for the corresponding $r$-matrix implies $\h_{ab}\x^a \x^b = 0$. However, a direct check shows that this is a solution to supergravity equations for any choice of $(\x_0,\x_1)$, i.e. a two-parametric family of solutions. Certainly, this is not a surprise since CYBE is a sufficient, rather than a necessary condition for a deformation to be a solution.

An interesting question here is that of the nature of the additional singularity at $z^3 = \x_ax^a$. Given the results of \cite{Barakin:2025jwp} we find it natural to think of these as not of a new horizon in the deformed $\AdS_3$, but as the region where dynamics of the string becomes non-relativistic. The intuition behind this expectation is that this singularity has the same origin as the critical electric field of \cite{Seiberg:2000gc,Seiberg:2000ms}. We find it interesting to check this explicitly by looking at spectrum of the string near this singularity.

\textbullet\;PP+PM-deformation is a combination of two deformation above. The bi-vector read
\begin{equation}
    \b = \fr12\g P_0 \wedge P_1 + \fr12 \x^a P_a \wedge M = \fr12 \left(\g + \x_a x^a \right)\dt_0 \wedge \dt_1.
\end{equation}
Applied to the near-horizon solution \eqref{eq:F1_NH} one gets the full string solution deformed by the PM deformation. Written using the radial $\AdS_3$ coordinate the solution is
\begin{equation}
    \begin{aligned}
        ds^2 & = \fr{\h_{ab}dx^a dx^b}{z^3 -\g- \x_a x^a} + \fr{R^2 dz^2}{4 z^3} + \fr{R^2}{z}d\W_7^2, \\
        B_2 & = \fr{dx^0 \wedge dx^1}{z^3 -\g- \x_a x^a}\\ 
        e^{-2\f}& = z^3 -\g- \x_a x^a
    \end{aligned}
\end{equation}
As before, no condition has to be imposed on $(\g,\x_0,\x_1)$ for the background to be a solution.

Performing the coordinate transform back to the radial coordinate $r^2 = R^2/z$ we obtain the solution of the same form as in \eqref{eq:F1m} with
\begin{equation}
        H(r) = -\g +\fr{R^6}{r^6} - \x_a x^a.
\end{equation}

\textbullet\;PP+PM+PD-deformation is non-abelian and inclu\-des a conformal Killing vector $D$. Although dilation is only a conformal Killing vector of the background, such a deformation produces a proper solution to supergravity equations. As we will see later this has nice colnnections to alternative representations of the flow corresponding to the PP+PM-deformation.

The bi-vector is given by
\begin{equation}
\begin{aligned}
    \b & = -\fr12 \g P_0\wedge P_1 + P_b \wedge \left[ \z^M_a \, \eta^{a b} M + \z^D_a \epsilon^{ab} D \right] \\
    &=-\fr12\g \dt_0\wedge \dt_1 -2\x_a x^a \, \dt_0 \wedge \dt_1  -2  \z_a \epsilon^{ab} \dt_b \wedge z \dt_z,
\end{aligned}
\end{equation}
where we denote $2\x=\z^M-\z^D$ and $\z=-\z^D$ for convenience. 
The resulting background is given by
\begin{equation}
    \begin{aligned}
        ds^2 = &\  \fr{1}{4z^3}e^{2\f}\Big[4z^3 \h_{ab}dx^adx^b - 4 R^2 z \z_a dx^a dz  + R^2 dz^2 (\g+z^3 + 4\x\cdot \s)\Big]+ \fr{R^2}{z}d\W_7, \\
        B_2  = &\ \fr{1}{z}e^{2\f}\Big( z dx^0\wedge dx^1 + \fr12\fr{R^2}{z} \e_{ab}\z^a dx^b \wedge dz\Big), \\
        e^{-2\f}  = &\ \fr{1}{z}\left(-R^2  \z^2 + z(\g+z^3 + 4\x\cdot \s)  \right)
    \end{aligned}
\end{equation}
where
\begin{equation}
    \begin{aligned}
        \z^2 &= \h_{ab}\z^a \z^b = -\z_0^2 + \z_1^2,\\
        \x\cdot \s & = \x_a x^a = \x_0x^0+\x_1x^1.
    \end{aligned}
\end{equation}
Although $D$ is not a Killing vector, this background appears to solve to supergravity equations upon the following condition
\begin{equation}
    \z^2 + 2\z\cdot \x = 0.
\end{equation}

As advertised earlier, this solution upon the coordinate transformation
\begin{equation}
     x^a = y^a + \fr{R^2}{2}\z^a \fr{1}{z},
\end{equation}
the background simplifies significantly and becomes
\begin{equation}
    \begin{aligned}
         ds^2 & = \fr{\h_{ab}dy^ady^b}{  \g+z^3 + 4(\x\cdot y)} + \fr{R^2 dz^2}{4z^3} + \fr{R^2}{z}d\W_7, \\
        B_2 & = \fr{ dy^0\wedge dy^1}{  z^3 + 4(\x\cdot y)}, \\
        e^{-2\f} & = \g+z^3 + 4(\x\cdot y),
    \end{aligned}
\end{equation}
that is precisely the PP+PM-deformed background. Going back to the $r$-coordinate one obtains the fundamental string solution \eqref{eq:F1m} with
\begin{equation}
    H(r) = \g+   4\x\cdot y + \fr{R^6}{r^6}.
\end{equation}

Now, it is worth mentioning that the bi-vector
\begin{equation}
    \b = \fr12 P_b \wedge \left[ \z^M_a \, \eta^{a b} M + \z^D_a \epsilon^{ab} D \right] 
\end{equation}
together with the condition $\z\cdot(\z+2\x)=0$ encodes deformations of six different classes. To see that it is convenient to introduce a vector $\bar{\z} = \z + 2\x$ to arrive at the orthogonality condition $\z\cdot \bar\z=0$, that has six different solutions:
\begin{equation}
\label{eq:class_zeta}
    \begin{aligned}
        & 1: && \z = (u,0), && \bar{\z} = (0,-v), && 2\x = (u,v), \\ 
        & 2: && \z = (0,-v), && \bar{\z} = (u,0), && 2\x = -(u,v), \\
        & 3: && \z = (0,0), && \bar{\z} = (-u,-v), && 2\x = (u,v),\\
        & 4: && \z = (u,v), && \bar{\z} = (0,0), && 2\x = (u,v),\\
        & 5: && \z = (u,u), && \bar{\z} = \l (u,u), && 2\x = (1-\l)(u,u), \\
        & 6: && \z = (u,-u), && \bar{\z} = \l(u,-u), && 2\x = (1-\l)(u,-u), 
    \end{aligned}
\end{equation}
where $u\neq v$ and $\l \in \mathbb R$. Vectors $\z$ and $\bar{\z}$ from different classes cannot be turned into each other by boosts in two dimensions. The corresponding bi-vectors (without the common $\g/2\dt_0\wedge \dt_1$ term) read
\begin{equation}
\label{eq:class_beta}
    \begin{aligned}
        \b_1 & = P_1\wedge (u D + v M), \\
        \b_2 & = P_0\wedge (v D + u M), \\
        \b_3 & = (u P_0  + v  P_1)\wedge M,\\
        \b_4 & = (-v P_0 + u  P_1)\wedge D ,\\
        \b_5 & = u (P_0 + P_1)\wedge (\l M + D) ,\\
        \b_6 & = u (P_0 - P_1)\wedge (\l M - D),
    \end{aligned}
\end{equation}
At the level of the background this information is partially lost since it depends only on $\x$ upon a proper coordinate transformation. Indeed, the all six classes give the same deformed background if one allows $u = v$ and avoids $\l=1$. The case $\l=1$ is special as in this case $\x=0$ and the deformation can be reverted by a coordinate transformation. From the algebraic point of view the point $\l=1$ is special as in this case the deformation is abelian
\begin{equation}
    [P_0\pm P_1, M\pm D] =0,
\end{equation}
We conclude that bi-vector deformations by $\b=u (P_0 \pm P_1) \wedge (M \pm D)$ are symmetries of the string background, even though $D$ is not an isometry. The bi-vectors listed in \eqref{eq:class_beta} are all simple in the classification provided in Appendix \ref{app:class}, and fall into classes $r_{1,\dots,6}$.

\subsection{String on deformed backgrounds}

All our deformed backgrounds considered above are of the form
\begin{equation}
    \begin{aligned}
        ds^2 & = H(r)^{-1} \h_{ab}dx^ad x^b  + g_{mn} dx^m dx^n \, \\
        B_2  & = H(r)^{-1} dx^0\wedge d x^1  + \fr12b_{mn} dx^m \wedge dx^n \,
    \end{aligned}
\end{equation}
where $a=0,1$ and $m,n=1,\dots,8$, and $H(r)$ is assumed to a function of deformation parameters. The Nambu--Goto string action, that we understand as the bosonic part of the full GS action, is given by 
\begin{equation} 
    \begin{aligned}
        S[X^{\m}](\r) & = - \int   d^2\s \sqrt{- \det G} - \int B_2  \\
    &=  - \int  d^2\s \left[ H^{-1} \left( \sqrt{|\det \left[ \eta_{\a \b} + m_{\a \b} |\right]} -1 \right)  -\fr{\e^{\a \b}}2 b_{\a \b} \right],
    \end{aligned} 
\end{equation}
where $\h_{\a\b}=\diag[-1,1]$ is the Minkowski metric tensor in 2D, we gauge fix $\s^{0,1} = x^{0,1}$, and denote
\begin{equation}
     m_{\a \b} \equiv H(r) g_{mn} \dt_\a x^m \dt_\b x^n.
\end{equation}
The standard pull-back of the Kalb--Ramond 2-form  in transverse direction is denoted by $b_{\a \b} = b_{mn} \dt_\a x^m \dt_\b x^n$. The energy-momentum tensor for this action is given by the following expression
\begin{equation}
\label{eq:T_PP}
\begin{aligned}
        T^\a{}_\b&= \d^\a{}_\b \mc{L} - \fr{\dt \mc{L}}{\dt\dt_\a x^m}\dt_\b x^m\\
        &=H^{-1}\left(\d^\a{}_\b  - \sqrt{-\det M}(M^{-1})^\a{}_\b\right),    
\end{aligned}
\end{equation}
where $M_{\a\b} = \h_{\a\b} + m_{\a\b}$ and the indices are raised and lowered by $\h_{\a\b}$. The inverse of the $2\times 2$ matrix $M_{\a\b}$ can be rewritten in a more convenient form  using
\begin{equation}
    \begin{aligned}   
        (M^{-1})^{\a\b} & = (\det M)^{-1} \e^{\a\g}\e^{\b\d}M_{\g\d}\\
        &= -2 (\det M)^{-1} \h^{\b[\a}\h^{\g]\d}(\h_{\g\d} + m_{\g\d}) \\
        & = (\det M)^{-1}\big(m^{\a\b} - \h^{\a\b} - \h^{\a\b}m^\g{}_\g\big),
    \end{aligned}
\end{equation}
where we have used the identity $\e^{\a\b}\e_{\g\d} = -2 \d^{[\a}{}_\g\d^{\b]}{}_{\d}$ with two raised indices. Substituting this back into \eqref{eq:T_PP} we obtain
\begin{equation} 
    \begin{aligned}
        T^\a{}_\b = &\ H(r)^{-1}(-\det M)^{-\fr12} \Big( m^\a{}_\b - \d^\a{}_\b \big( 1 + m_\g{}^\g -\sqrt{-\det M}\big) \Big),
    \end{aligned} 
\end{equation}
where the indices are raised and lowered by the flat metric $\h_{\a\b}$. Now, understanding the function $H$ as a function $H=H(\g)$ of a deformation parameter $\g$ it is straightforward to derive the following general relation
\begin{equation}
\label{eq:def_gen}
    \fr{\dt S(\g)}{\dt \g} = -\fr12\int d^2 \s \dt_\g H(\g) \det T^\a{}_\b.
\end{equation}

Let us now turn to various examples

\textbullet\;\textbf{Abelian flow} (PP-deformation).  In this case the deformed harmonic function is given by
\begin{equation}
    H(r) = \r + \fr{R^6}{r^6}\,,
\end{equation}
and the flow equation is simply
\begin{equation}
    \dt_\r S(\r)  = -\fr12\int d^2\s \det T,
\end{equation}
that is the standard $\mathrm{T}\bar{\mathrm{T}}$ flow equation.  It is suggestive to investigate various limits of the deformation parameter.

For $\r \to \infty$ the background space-time becomes approximately flat, that can be understood as probing the region far away of the stack of fundamental strings sitting at $r=0$. Taking the sedimentation point of view advocated in \cite{Barakin:2025jwp} one may understand this limit as removing the string charge from the background, that naturally leads to the flat space-time. To have a proper limit one must rescale the longitudinal coordinates as $\bar\s^a = \r^{\fr12}\s^a$. The string action then becomes simply the flat Nambu--Goto action with a pure gauge B-field
\begin{equation}
\label{eq:pp_flat}
    S^{(0)}_{\r\to \infty} = - \int d^2 \bar\s \Big(\sqrt{1+\tilde{m}^\a{}_\a + \det \tilde{m}^\a{}_\b}-1\Big),
\end{equation}
where $\tilde{m}_{\a\b} = \dt_\a x^m \dt_\b x^n \d_{mn}$. Doing so, however, we lose completely the information about the core string charge we started with, and hence it is impossible to recover the string action on the F1 background from \eqref{eq:pp_flat}. The same was observed in \cite{Barakin:2025jwp} as a statement, that PP-deformation of the flat space-time does not give the string background.

It is then natural to keep the first non-vanishing term in the $\r \to \infty$ limit, that gives up to the first order
\begin{equation}
\begin{aligned}
    &S^{(1)}_{\r\to \infty} = - \int d^2 \s \Big[\big(\sqrt{\l}-1\big) + \fr{\r^{-1} R^6}{r^6}\Big(-\sqrt{\l}+1 + \fr12 \fr{\tilde{m}^\a{}_\a + 2 \det \tilde{m}^\a{}_\b}{\sqrt{\l}}\Big)\Big],
\end{aligned}
\end{equation}
where we denote $\l =1+\tilde{m}^\a{}_\a + \det \tilde{m}^\a{}_\b$ for compactness of notations. Rather expected is that the term linear in $\r^{-1}$ is nothing but the trace of the energy-momentum tensor for the action related to the $S^{(0)}_{\r\to \infty}$, and we may write
\begin{equation}
    S^{(1)}_{\r\to \infty}=S^{(0)}_{\r\to \infty} + \fr12 \r^{-1} \int d^2\s \,\fr{R^6}{r^6}\,T^\a{}_\a.
\end{equation}
This action possesses information of what would be the core string charge. It is tempting to declare that deformation by such a composite operator that depends on the field $r(\s)$ (similarly to vertex operators) gives the flow from NG string on flat background to the NG string on the string background. However, there is a subtlety at small values of the background string charge, i.e. in the beginning of such a flow, where the probe string approximation breaks down. In other words, we cannot analyse emergence of the core string charge in the probe string approximation. This might be the fundamental reason behind the fact that the full fundamental string background cannot be obtained by a bi-vector deformation of the flat space-time.

Another possibility is to take the same limit $\r \to \infty$ but with a different rescaling, that allows to end up at the sector of non-relativistic strings. For that we rescale $\bar\s^a = \r \s^a$, that gives the background
\begin{equation}
\label{eq:GO}
    \begin{aligned}
        ds^2 & = \r\, \h_{ab}d\s^a d\s^b + ds_\perp^2 , \\
        B & = \r\, d\s^0 \wedge d\s^1.
    \end{aligned}
\end{equation}
In this form it is the relativistic Gomis--Ooguri background, however one should keep in mind the limit $\r \to \infty$ that makes \eqref{eq:GO} non-relativistic. The string action in this limit becomes
\begin{equation}
    S = -\int d^2 \s \left[\r \sqrt{1 + \r^{-1}\tilde{m}^\a{}_\a + \r^{-2}\det \tilde{m} } - 1\right],
\end{equation}
that completely reproduces the whole story of \cite{Blair:2020ops}. We conclude, that $\rho \to \infty$ together with the rescaling as above should be understood as the correct way of taking the non-relativistic limit of the fundamental string on the fundamental string background. 

It is worth mentioning that here in contrast to a space-time bi-vector deformation of the flat space-time, where a clear notion of the critical deformation exist, for finite deformations the harmonic function degenerates at certain values of $r$. Requiring $H(r_{cr})=0$ one may introduce the notion of critical distance $r_{cr}$ from the core that correlates with a certain value of deformation $\r_{cr}$ as
\begin{equation}
    \r_{cr} + \fr{R^6}{r^6_{cr}}=0.
\end{equation}
It is tempting to speculate that near such defined $r_{cr}$ the string spectrum becomes non-relativistic, pretty much as that of the string on a critical deformation ( with $\r_{cr}=-1$) of the flat space-time. Such an analysis stands beyond the scope of the present paper, and we hope to return to it in an upcoming publication.

\textbullet\;\textbf{Non-abelian flow} (PP+PM-deformation). The harmonic function is given by
\begin{equation}
    H(r) = \g + \x_a x^a + \fr{R^6}{r^6},
\end{equation}
and the deformation parameters $\x_a$ and $\g$ are not restricted by further constraints. In these coordinates the flow equation takes the following form
\begin{equation}
    \fr{\dt S}{\dt \x_a} = 4 \int d^2\s y^a \det T^\a{}_\b.
\end{equation}
To relate this to the boost and momentum currents, as suggested by the form of the deformation, we turn to light-cone coordinates on the world-volume as
\begin{equation}
    x^\pm = x^0 \pm x^1, \quad ds^2 = - dx^+ dx^-.
\end{equation}
The Noether boost current $j_B^\a$ in 2D is defined as follows
\begin{equation}
    M^{\a,\b\g} = x^\b T^{\a\g} - x^\g T^{\a\b} = j_B^\a \e^{\b\g},
\end{equation}
where we do not include the spin part since the world-volume fields $x^m$ are scalars. In the light-cone coordinates the boost current 1-form $j_B = j_{B\a}dx^\a$ and the momentum current $T^\a = T^\a{}_\b dx^\b$ takes the following form
\begin{equation}
    \begin{aligned}
        j_B &= (x^+ T_{++} + x^- T_{+-})dx^+ + (x^+ T_{+-} + x^- T_{--})dx^-, \\
        T^+ & = T_{++}dx^+ + T_{+-}dx^-, \\
        T^- & = T_{+-}dx^+ + T_{--}dx^-.
    \end{aligned}
\end{equation}
Now, observing the identity
\begin{equation}
    j_B \wedge T^\pm = x^\pm \det T^\a{}_\b dx^+ \wedge dx^-,
\end{equation}
we write the non-abelian PM-flow as follows
\begin{equation}
    \fr{\dt S}{\dt \x_\pm} = \int j_B \wedge T^\pm.
\end{equation}
This form is a complete analogue of that for the $\mathrm{T}\bar{\mathrm{T}}$ flow, written in real light-cone coordinates:
\begin{equation}
    \dt_\g S = \int T^+ \wedge T^-.
\end{equation}

We find it interesting to also consider the dilation current $j_D^\a = x_\b T^{\a\b}$, that in the light-cone coordinates can be written as the following 1-form
\begin{equation}
    j_D = (x^+ T_{++} - x^- T_{+-})dx^+ + (x^+ T_{+-} - x^- T_{--})dx^-.
\end{equation}
Its wedge product with the momentum currents gives
\begin{equation}
    j_D \wedge T^\pm = \mp x^\mp \det T^\a{}_\b dx^+ \wedge dx^-.
\end{equation}
Therefore, the same PM-flow can be written as a $DT$-deformation as
\begin{equation}
    \fr{\dt S}{\dt \x_\pm} = \pm \int j_D \wedge T^\mp.
\end{equation}
Certainly, any combination of $DT$ and $MT$ flows is also possible. This is a reflection of the fact, that the PM+PD-deformation (with certain additional constraints) is the same as the PM-deformation up to a coordinate transformation.

It is worth mentioning that although the obtained flows for non-abelian background deformations look like the familiar current-current flows, the corresponding 1-forms $j_B$ and $j_D$ are not conserved in the 2D QFT \footnote{We are grateful to the anonymous referee for point out this issue.}. The reason is the simple fact that non-abelian deformations typically destroy the isometries they are built from. In this case, the deformed background no longer has the boost isometry, and divergences of the currents are explicitly proportional to $\x^\alpha$.

\subsection{The string on \texorpdfstring{AdS$_3$}{AdS3}}

We consider now the case of a background where the algebra $\frso(2,2)$ is an isometry algebra. Therefore classical Yang--Baxter equation together with the unimodularity conditions are sufficient for a deformation to generate a solution. The background is $\AdS_3\times \SS^3 \times \TT^4$ and is explicitly written as
\begin{equation}
\begin{aligned}
    ds^2   = &\  \frac{R^2}{z^2} 
    \left(- d x_0^2+ d x_1^2  + dz^2 \right) + R^2 \Big( d\chi^2 +\sin^2\chi
    \left( d\theta^2+\sin^2\theta\, d\varphi^2
    \right) \Big)     + ds_{\mathbb{T}^4}^2, \\
    B_2  = &  -\frac{R^2}{z^2}dx^0\wedge dx^1 + 
    \fr{R^2}{2}
    \left(
        \chi-\frac12\sin 2\chi
    \right)
    \sin\theta d\q \wedge d\f.
\end{aligned}
\end{equation}
The background has constant dilaton and is supported by the 3-form NS-NS flux $H_3 = vol_{\AdS_3} + vol_{\SS^3}$. The deformation is 
\begin{equation}
    \b = \fr14 \g (P_0 + P_1)\wedge (K_0 + K_1),
\end{equation}
that belongs to the class $r_8$ of Appendix \ref{app:class}. Using the explicit form of special conformal generator $K_a = x^2 P_\a + 2 x_\a D$ we obtain the following deformation bi-vector as
\begin{equation}
    \b = \g (x^-)^2 \dt_+ \wedge \dt_- + \g x^- z \dt_+ \wedge \dt_z.
\end{equation}
It proves convenient to use light-cone coordinates $ x^\pm = x^0 \pm x^1$. The deformed background is given by
\begin{equation}
    \begin{aligned}
        ds^2  = & \ \fr{R^2}{z^2}dz^2 - \fr{R^2}{z^2 + \g (x^-)^2} dx^-\left(dx^+ + \g \fr{dz}{z}x^-\right) + ds_{\SS^3}^2 + ds_{\TT^4}^2,\\
        B_2  = & -\fr{R^2}{2}\fr{1}{z^2 + \g (x^-)^2}\left(dx^+ + \g \fr{dz}{z}x^-\right)\wedge dx^- + b_\perp,\\
        e^{-2\f}  =&\ 1 + \g \left(\fr{x^-}{z}\right)^2,
    \end{aligned}
\end{equation}
where $b_\perp$ stands for the 2-form components in directions transverse to $\AdS_3$. After the coordinate transformation
\begin{equation}
    x^+ \to x^+ - \g x^- \log z,
\end{equation}
and a gauge transformation of $B_2$ the resulting solution reads
\begin{equation}
    \begin{aligned}
        ds^2  = & \fr{R^2}{z^2}dz^2 - \fr{R^2}{z^2 + \g (x^-)^2} dx^-\left(dx^+ -\g \log z dx^-\right) + ds_{\SS^3}^2 + ds_{\TT^4}^2,\\
        B_2  =& -\fr{R^2}{2}\fr{1}{z^2 + \g (x^-)^2}dx^+\wedge dx^- + b_\perp,\\
        e^{-2\f}  = &\ 1 + \g \left(\fr{x^-}{z}\right)^2.
    \end{aligned}
\end{equation}

The metric cannot be brought into the diagonal form as we did for deformations of the string background, however, derivation of the flow equation in this case works pretty similarly. The string action on the deformed background is given by
\begin{equation}
    S = - \int d\x^+ d\x^- \left(\sqrt{-\det \bar{M}_{\a\b}}  + B_{+-} -\fr12 \e^{\a\b}b_{\a\b}\right),
\end{equation}
where
\begin{equation}
    \begin{aligned}
        \bar{M}_{\a\b} & = g_{\a\b} + \bar{m}_{\a\b}, \\
        \bar{m}_{\a\b} & = g_{mn}\dt_\a x^m \dt_\b x^n,\\
        \bar{b}_{\a\b} & = B_{mn}\dt_\a x^m \dt_\b x^n.
    \end{aligned}
\end{equation}
Here the indices $\a\b = +,-$ and the indices $m,n$ run the remaining directions, and as before we made the gauge choice $\x^\pm = x^\pm$. The energy-momentum tensor is given by
\begin{equation}
\begin{aligned}
    T^\a{}_\b  = & -\left(\d^\a{}_\b B_{+-}+ \sqrt{-\det\bar{M}_{\a\b}}(\bar{M}^{-1})^{\a\g}g_{\g\b}\right)\\
     = &\ \d^\a{}_\b\left(-B_{+-} - g\sqrt{-\det \bar{M }}(1+\bar{m}^\g{}_\g)\right) + g \sqrt{-\det \bar{M }} \bar{m}^\a{}_\b,
\end{aligned}
\end{equation}
where indices are lowered and raised by $g_{\a\b}$ and $g = -\det g_{\a\b}$.

Now, derivative of the action along the deformation parameter is given by
\begin{equation}
    \begin{aligned}
        \fr{\dt S}{\dt \g}  & =\fr14 \int d^2\x \fr{-(x^-)^2}{z^2 + \g (x^-)^2}\left[g\sqrt{|\det \bar{M }|}(2+\bar{m}^\a{}_\a ) - 2 B_{+-}\right] \\
        & = \fr{1}{4}\int d\x^+ d\x^- \left(\fr{x^-}{R}\right)^2 \det T = \fr1{4R^2} \int d\x^+ d\x^- j_{K_+} \wedge T_+.
    \end{aligned}
\end{equation}
We conclude, that the $P_+\wedge K_+$ deformation of the $\AdS_3\times \SS^3 \times \TT^4$ background induces the flow by the $K_+\wedge T_+$ operator on the 2D string world-volume.

\subsection{D0-brane and uni-vector deformations}
\label{sec:D0}

The same formalism can be applied to Dp-branes and the corresponding $p+1$-vector deformations. We start with a D0-brane on a uni-vector deformed D0-brane background. The undeformed solution reads
\begin{equation}
\label{eq:D0}
    \begin{aligned}
        ds^2 & = - H^{-\fr12}dt^2 + H^{\fr12}(dr^2 + r^2 d\W^2) ,\\
        C_1& = g_s^{-1} H^{-1}dt, \\
        e^{\f} & = g_s H^{\fr34}
    \end{aligned}
\end{equation}
has only one Killing vector $\dt_t$ along the world-line. A uni-vector deformation generated by $\a \dt_t$ along this Killing vector is the precise analogue of the $P_0\wedge P_1$ bi-vector deformation, that corresponds to the $\mathrm{T}\bar{\mathrm{T}}$ flow. If one performs no further rescaling
of the time coordinate $t$ as in \cite{Barakin:2026mxz}, the deformed background takes the same form as in \eqref{eq:D0} but with a different harmonic function
\begin{equation}
    H \to 2\g + H.
\end{equation}

It is then suggestive to repeat the whole procedure of deformation of the D0-brane action of the background \eqref{eq:D0} as we did it for the fundamental string on its own background. The D0-brane action reads
\begin{equation}
    S_{D0} = - T_0 \int d\t e^{-\f}\sqrt{-(G+B)_{\t\t}} + T_0 \int C_1,
\end{equation}
where $G_{\t\t} = g_{\m\n}\dt_\t X^\m\dt_\t X^\n$ is the pull-back of the space-time metric to the world-line, and the same for $B_{\t\t}$. On the background \eqref{eq:D0} the action takes the following form
\begin{equation}
    S_{D0} = -T_0 \int d\t H^{-1}\left(\sqrt{1 - H \dt_0 X^m \dt_0 X^m}-1 \right),
\end{equation}
where we fixed the gauge freedom by choosing $X^0 = \t$, and the index $m$ runs all transverse directions. The energy-momentum tensor on the world-line has only one component and reads
\begin{equation}
    T = T_0 H^{-1}\left[(1 - H \dt_0 X^m \dt_0 X^m)^{-\fr12} - 1 \right].
\end{equation}
Let us now write the flow equation corresponding to this uni-vector deformation. Taking derivative of the action along $\g$ and doing some simple algebra gives
\begin{equation}
\begin{aligned}
    \dt_\g S_{D0}    &=-\fr{T_0}{2} \int d\t H^{-2} \left(2 + \fr{-2 + H \dt_0 X^m \dt_0 X^m}{ \sqrt{1 - H \dt_0 X^m \dt_0 X^m}} \right) \dt_\g H\\
    &= \fr{1}{2}  \int d\t \dt_\g H \fr{T^2 }{T_0 + H T},
\end{aligned}
\end{equation}
that is in consistency with the $p=1$ flow of \cite{Blair:2024aqz} (again up to the extra factor of the harmonic function). 

An interesting observation is that this expression simplifies significantly for ultra-relativistic D0-brane. Indeed, in this case the interval is almost light-like $ds \simeq 0$, that gives
\begin{equation}
    1 - H \dt_0X^m \dt_0 X^m \simeq 0.
\end{equation}
The energy $T$ then satisfies $T_0 \ll H T$ and the flow equation becomes approximately
\begin{equation}
    \dt_\g S = \fr12 \int d\t H^{-1} \dt_\g H \, T. 
\end{equation}

\subsection{D3-brane and quadri-vector deformations}

As it was shown in \cite{Barakin:2025jwp} the D3-brane background transforms under quadri-vector deformations similarly to the fundamental string background under bi-vector deformation. In particular, one observes sedimentation, when under the deformation 
\begin{equation}
    \W_4 = \g P_0\wedge P_1\wedge P_2\wedge P_3
\end{equation}
the harmonic function simply shifts as $H(r) \to \g + H(r)$. It is therefore natural to extend the results above to the case of a D3-brane probe on a deformed D3-brane background. Start with the undeformed background
\begin{equation}
    \begin{aligned}
        ds^2 & = H^{-\fr12}ds_{||}^2 + H^{\fr12}(dr^2 + r^2 d\W_5^2), \\
        C_4 & = H^{-1}dx^0\wedge dx^1\wedge dx^2\wedge dx^3,
    \end{aligned}
\end{equation}
with
\begin{equation}
    H(r) = 1 + \fr{r_0^4}{r^4}.
\end{equation}
In the absence of world-volume gauge fields the gauge fixed D3-brane action on this background is given by
\begin{equation}
    S_{D3} = - T_{D3} \int d^4 \s H^{-1}\left(\sqrt{-\det M_{\a\b}}-1\right),
\end{equation}
where we define
\begin{equation}
    M_{\a\b} = \h_{\a\b} + H(r) \d_{mn} \dt_\a X^m \dt_\b X^n.
\end{equation}
Assuming that the harmonic function depends on a deformation parameter $\g$ we obtain for derivative of the action
\begin{equation}
\label{eq:parS_D3}
\begin{aligned}
    \dt_\g S     &=- T_{D3} \int d^4 \s H^{-2} \dt_\g H\Big[1 + \sqrt{-\det M_{\a\b}} - \fr12 M^{\a\b}\h_{\a\b}\Big].
\end{aligned}
\end{equation}
The energy-momentum tensor has the form 
\begin{equation}
    T^\a{}_\b = H^{-1}(\d^\a{}_\b + \t^\a{}_\b),
\end{equation}
where it proves convenient to introduce 
\begin{equation}
    \t^\a{}_\b = - \sqrt{-\det M}(M^{-1})^\a{}_\b.
\end{equation}
Here comes the subtlety of the D3-brane case, that arises from the relation $\det \t = \det M$. Indeed, since $\det T$ is a polynomial of powers of $\Tr\,\t^n$ with integer $n$, one in general cannot get rid of the square root in \eqref{eq:parS_D3}. It is however possible to rewrite the derivative \eqref{eq:parS_D3} in a form close to the standard $\mathrm{T}\bar{\mathrm{T}}$ expression, if an additional condition on $T^\a{}_\b$ is imposed. Before doing so, consider the full expression and rewrite it in terms of the energy-momentum tensor.

For that we introduce short-hand notations for traces $t_k = H^k \Tr T^k$, and the elementary symmetric polynomials
\begin{equation}
    \begin{aligned}
        e_1 & = t_1, \\
        e_2 & = \fr12 (t_1^2 - t_2),\\
        e_3& = \fr16 (t_1^3 - 3 t_1 t_2 + 2 t_3)\\
        e_4 & = \fr1{24}(t_1^4 - 6 t_1^2 t_2 + 3 t_2^2 + 8 t_1 t_3 - 6 t_4),
    \end{aligned}
\end{equation}
that are related to eigenvalues $\l_i$ of the matrix $H(r)T^\a{}_\b$ as follows
\begin{equation}
    \begin{aligned}
         & e_1 = \sum_i \l_i, && e_2 = \sum_{i<j}\l_i\l_j, \\
         & e_3 = \sum_{i<j<k}\l_i\l_j\l_k, && e_4 = \l_1\l_2\l_3\l_4.
    \end{aligned}
\end{equation}
It is then  straightforward to express $\det\t = \det(H T - \mathbf{1})$ in terms of $e_k$ as follows 
\begin{equation}
    \begin{aligned}
        \dt_\g S_{D3} = - T_{D3} & \int d^4 \s  H^{-2}\dt_\g H \Big( \fr12 e_1 - 1 +\sqrt{ 1 - e_1 + e_2 - e_3 + e_4} \Big).
    \end{aligned}
\end{equation}

As we mentioned, one cannot get rid of the square root in the general case, however in this form the expression suggest an additional constraint to be imposed such that to make it possible. Indeed, if the matrix $T^\a{}_\b$ (equivalently, the matrix $\t$) has eigenvalues that coincide pairwise, then $\det\t$ becomes a full square. To write this constraint explicitly we need expressions for eigenvalues $\l_i$ in terms of the quantities $e_i$. For that one usually starts with the characteristic equation
\begin{equation}
    \l^4 - e_1 \l^3 + e_2 \l^2 - e_3 \l  + e_4 =0.
\end{equation}
The substitution $\l = y+1/4 e_1$ turns this equation into 
\begin{equation}
    y^4 + py^2 + q y + r =0,
\end{equation}
where
\begin{equation}
    \begin{aligned}
        p &= e_2-\frac{3}{8} e_1^2, \\
        q &= -\frac18 e_1^3+\frac12 e_1 e_2- e_3,\\
        r & =-\frac{3}{256} e_1^4+\frac{1}{16} e_1^2 e_2-\frac{1}{4}e_1 e_3+e_4.
    \end{aligned}
\end{equation}
This equation has four roots
\begin{equation}
\label{eq:roots}
    \begin{aligned}
        y = \e \sqrt{z} + \h \sqrt{-2 p - z - \e \fr{2 q}{\sqrt{z}}},
    \end{aligned}
\end{equation}
where $\e^2 = \h^2 =1$ and $z$ is any non-zero root of the cubic equation
\begin{equation}
    z^3+2 p z^2 + (p^2-4r)z-q^2= 0.
\end{equation}
It is easy to see, that given $z\neq 0$ the only way to fuse two roots is to require that the square roots in \eqref{eq:roots} vanish. That together with the cubic equation on $z$ gives two constraints
\begin{equation}
\label{eq:conditions}
    \begin{aligned}
        q = 0 , \quad p^2 - 4 r = 0.
    \end{aligned}
\end{equation}
Using this condition it is straightforward to rewrite the expression in the square root as follows
\begin{equation}
    1 - e_1 + e_2 - e_3 + e_4 = \left(1 - \fr12 e_1 + \fr12 e_2 - \fr18 e_1^2\right)^2.
\end{equation}
Generically, taking square root would give two branches, and one therefore is interested in the sign of the expression in parentheses. For the case in question it can be rewritten as follows
\begin{equation}
    \begin{aligned}
        1 - \fr12 e_1 + \fr12 e_2 - \fr18 e_1^2 & = \fr18 (\Tr \t)^2 - \fr14 \Tr \t^2 \\
        &=(H \l_1 - 1)(H \l_2 - 1) \\
        &= \fr1{16}\left(H(e_1 + 2 \sqrt{-2p})+1\right) \left(H(e_1 - 2 \sqrt{-2p})+1\right), 
    \end{aligned}
\end{equation}
where $p$ certainly must be negative. Now, we would like to determine sign of this expression, that is continuously connected to the undeformed limit. For that we also take the asymptotically flat region of the background, that is $r \to \infty$ and take the specific point-like D3-brane configuration. In this case $M_{\a\b}=\h_{\a\b}$ and $H(r) = 1$, that gives the positive sign. Now, going closer to the core of the background and by exciting the probe D3-brane such that it does not change the background much, one can not change the sign of this expression. Indeed, for that it would necessarily pass through zero, that will give vanishing energy-momentum tensor and a degenerate configuration. With this sign choice we are able to write the derivative \eqref{eq:parS_D3} in the following simple form
\begin{equation}
    \begin{aligned}
        \dt_\g S & = - \fr18  T_{D3} \int d^4 \s H^{-2}\dt_\g H \big(t_1^2 - 2 t_2\big) \\
         & = - \fr18  T_{D3} \int d^4 \s \dt_\g H \Big((\Tr T)^2 - 2 \Tr T^2\Big).
    \end{aligned}
\end{equation}
Substituting explicit dependence of the harmonic function $H(r)$ on the deformation parameter, one obtains flow equations for the D3-brane for various quadri-vector deformations. 

The conditions \eqref{eq:conditions} restrict the energy-momentum tensor of the D3-brane to be of 2+2-type, that is its eigenvalues coincide pairwise. From the world-volume point of view this means, that the brane configuration preserves a symmetry of the form $\rmSO(1,1)\times \rmSO(2)$ and has two principal stresses: along a 1+1-dimensional world-volume and along a two-dimensional plane. Such world-volume symmetry breaking is natural for bound state configurations such as D3-F1 or D3-D1, however our system does not have  world-volume gauge fields explicitly, and quadri-vector deformations adds a D3-brane charge rather than a D1 or an F1. One naturally may consider D3-branes wrapping two toroidal directions and write the corresponding scalar field configuration. Closer analysis of such systems stands beyond the scope of the present paper, therefore we leave this as an interesting observation deserving further research.

\section{The 11D membrane}
\label{sec:M2}

In this section we use the formalism of tri-vectors to derive flow equations for the membrane for abelian and non-abelian deformations.

\subsection{Deformations of the M2 background}

Bi-vector and quadri-vector deformations of 10-dimensional backgrounds considered above have natural generalization in 11 dimensions suggested in \cite{Bakhmatov:2019dow}. Such deformations are generated by a tri-vector parameter $\W^{mnk}$, that is taken in the tri-Killing ansatz
\begin{equation}
    \W^{mnk} = \fr1{4!}\r^{ABC}k_A^mk_B^n k_C^k,
\end{equation}
where $k_A^m$ are Killing vectors of the initial solution. The general formalism of such deformation has been developed in \cite{Gubarev:2020ydf}, while here we will be interested in the deformation found in \cite{Bakhmatov:2020kul} and dubbed PPP, PPM and DPP. These are tri-vector deformations of the $\AdS_4\times \SS^7$ solution to 11D supergravity equations, that can be understood as the near-horizon limit of the M2 background. The full M2-brane solution reads 
\begin{equation} 
\label{eq:M2_sol}
\begin{aligned}
    ds^2 & = H^{-\fr23} (-dx_0^2 + dx_1^2 + dx_2^2)  + H^{\fr13} \left( dr^2 + r^2 d\W_7^2 \right) \, \\
    C_3 & = H^{-1} d x^0\wedge dx^1 \wedge dx^2 \\
    H & = 1 + \frac{R^6}{r^6}.
\end{aligned} 
\end{equation}
For tri-vector deformations we again use the set of Killing vectors of the $\AdS_4$ part of the near-horizon region of the full solution, including the dilation Killing vector.  Explicitly the Killing vectors are given by
\begin{equation}
    \begin{aligned}
        P_a &= \dt_a, \quad M_{ab} = x_a\dt_b - x_b \dt_a, \\
        D& = - x^a \dt_a - z\dt_z,
    \end{aligned}
\end{equation}
where $a=0,1,2$ and $z$ denotes the radial coordinate in $\AdS_4$ defined as $z=R^2/r^2 $.

\textbullet\;PPP-deformation of $\AdS_4\times \SS^7$ is abelian and is generated by the following tri-vector
\begin{equation}
    \fr{R^3}4 \W = \g  \, P_0 \wedge P_1 \wedge P_2 = \g  \, \dt_0 \wedge \dt_1 \wedge \dt_2,
\end{equation}
The deformed background is of the same form as \eqref{eq:M2_sol} with the harmonic function
\begin{equation}
     H = \g + \frac{R^6}{r^6},
\end{equation}
and solves 11D supergravity equations for any value of $\g$.

\textbullet\;PPM+PPD-deformation is a combination of the PPM and DPP deformations found in \cite{Bakhmatov:2020kul}. The tri-vector is given by
\begin{equation}
    \begin{aligned}
        \fr{4}{R^3}\W & = \fr14 \z_M^a\e^{bcd}P_a\wedge P_b\wedge M_{cd} + \fr12 \z_{D\, a}\e^{abc}D\wedge P_b\wedge P_c \\
        & = \bar\z_a x^a \dt_0\wedge \dt_1\wedge \dt_1 - \fr z2 \z_{a}\e^{abc}\dt_b\wedge \dt_c\wedge \dt_z.
    \end{aligned}
\end{equation}
where we define $\bar\z = \z_M+\z_D$ and $\z = \z_D$.
This deformation generates a solution to 11D supergravity equations given the constraint
\begin{equation}
    \z^2 + 2 (\z\cdot \bar\z)=0,
\end{equation}
where $\z\cdot \bar\z = \z_a \bar\z^a$. The solution can be brought to the same form \eqref{eq:M2_sol} via the coordinate transformation $x^a \to x^a - 1/2\z^a z^{-1}$ with the defining function
\begin{equation}
    H = \bar\z_a x^a + \fr{R^6}{r^6}.
\end{equation}

\textbullet\;PPP+PPM+PPD-deformation is obtained from the previous one by adding the PPP-term. The background is again of the same form \eqref{eq:M2_sol} with the defining function
\begin{equation}
    H = \g+\bar\z_a x^a + \fr{R^6}{r^6}.
\end{equation}

\subsection{The membrane flow equation }

As in the Type II case, all our deformed 11D backgrounds are of the form
\begin{equation} 
\begin{aligned}
    ds^2 & =  H(r)^{-\fr23} ds_{||}^2  + H(r)^{\fr13} \left( dr^2 + r^2 d\W_7^2 \right) \, \\
    C_3 & = H(r)^{-1} dx^0\wedge dx^1 \wedge dx^2.
\end{aligned}
\end{equation}
The M2 action on this background with vanishing world-volume gauge fields is given by 
\begin{equation} 
\begin{aligned}
    S[X^{\m}] & = - T_{M2}\int  d^3\s \sqrt{- \det G} +T_{M2} \int C_3  = \\
    & = - T_{M2} \int d^3\s \, H(r)^{-1} \left( \sqrt{- \det M_{\a\b}} - 1 \right) \, , \\
    \end{aligned}
\end{equation}
where as before we denote $M_{\a\b}= \eta_{\a \b} + m_{\a \b} $ with
\begin{equation}
    \begin{aligned}      
     & m_{\a \b} = H(r) \d_{m n} \dt_\a X^m \dt_\b X^n
\end{aligned} 
\end{equation}
The energy-momentum tensor for this action is given by
\begin{equation} 
    \begin{aligned}
        T^\a{}_\b = & H(r)^{-1} \left(\d^\a{}_\b - \sqrt{-\det M} (M^{-1})^\a{}_\b \right)
    \end{aligned} 
\end{equation}

Assuming that the harmonic function depends on a deformation parameter $\g$ we obtain the following for derivative of the action
\begin{equation}
\label{eq:parS_M2}
\begin{aligned}
    \dt_\g S    &= T_{M2}\int d^3 \s \dt_\g H^{-1} \left[1 + \fr12\sqrt{-\det M}\left(1 - M^{\a\b}\h_{\a\b}\right)\right],
\end{aligned}
\end{equation}
where $M^{\a\b} = (M^{-1})^\a{}_\d\h^{\b\d}$. As in the case of the fundamental string it is possible to rewrite this equation in terms of $\det T^\a{}_\b$ and trace expressions. For that as in the D3-brane case we introduce
\begin{equation}
    \t^\a{}_\b = -\sqrt{-\det M}(M^{-1})^\a{}_\b, 
\end{equation}
that gives
\begin{equation}
    \begin{aligned}
        \det \t & = \sqrt{-\det M}, \\
        \Tr \, \t & = -\sqrt{-\det M}M^{\a\b}\h_{\a\b}.
    \end{aligned}
\end{equation}
On the other hand $\det \t$ can be expressed in terms of $\det T$ as follows
\begin{equation}
\label{eq:dettau_M2}
    \begin{aligned}
        \det \t = H^3 \det T - 1 - \fr12 (\Tr\,\t)^2 + \fr12 (\Tr\, \t^2) - \Tr\,\t
    \end{aligned}
\end{equation}
Similarly we have
\begin{equation}
\label{eq:TrT_M2}
    \begin{aligned}
        \Tr\,T & = H^{-1}(3 + \Tr\,\t),\\
        \Tr\,T^2 & = H^{-2}(3 + 2 \Tr\,\t + \Tr\,\t^2).
    \end{aligned}
\end{equation}
Substituting \eqref{eq:dettau_M2} into \eqref{eq:parS_M2} and taking into account \eqref{eq:TrT_M2} we obtain
\begin{equation}
\begin{aligned}
    \dt_\g S_{M2}     &= - \fr12T_{M2} \int d^3 \s \dt_\g H \left[\fr12 (\Tr\, T)^2 - \fr12 \Tr\,T^2 - H(r)\det T\right].
\end{aligned}
\end{equation}
The first two terms reproduce the similar result for the fundamental string, while in full we get the same expression as in \cite{Blair:2024aqz} but with additional dependence on $r$ inside the harmonic function.

Let us now consider examples

\textbullet\;\textbf{Abelian flow} (PPP-deformation). Derivative of the action along the deformation parameter $\g$ reads
\begin{equation}
\begin{aligned}
    \dt_\g S_{M2} 
    &=- \fr12T_{M2} \int d^3 \s  \left[\fr12 (\Tr\, T)^2 - \fr12 \Tr\,T^2 - H(r)\det T\right].
\end{aligned}
\end{equation}
This can be rewritten in the form similar to the standard representation of the $\mathrm{T}\bar{\mathrm{T}}$ deformation upon introducing a current $I^a = d\s^a$ trivially conserved on the flat M2 world-volume. Denoting the momentum current as before $T^a = T^a{}_b d\s^b$ we write the flow equation as
\begin{equation}
\begin{aligned}
    \dt_\g S   &=-\fr12 T_{M2} \int \left(\fr12 \e_{abc}T^a\wedge T^b \wedge I^c - H(r)\, T^0\wedge T^1 \wedge T^2\right).
\end{aligned}
\end{equation}

\textbullet\;\textbf{Non-abelian flow} (PPM+PPD-deformation). The defining function in this case is
\begin{equation}
    H = \bar{\z}_a x^a + \fr{R^6}{r^6},
\end{equation}
and derivative of the action reads 
\begin{equation}
\begin{aligned}
    \fr{\dt S}{\dt \z_a}   &=-\fr12 T_{M2}  \int d^3 \s  \s^a \left[\fr12 (\Tr\, T)^2 - \fr12 \Tr\,T^2 - H(r)\det T\right].
\end{aligned}
\end{equation}
To write this as a wedge product if three currents we will need yet another trivially conserved current, that is $K^{ab} = \s^{[a}d\s^{b]}$. The flow equation is then
\begin{equation}
    \begin{aligned}
        \fr{\dt S}{\dt \z_a} =-\fr12 T_{M2}\int\bigg( \e_{bcd}T^b\wedge T^c \wedge K^{ab} - \fr12 \e_{bcd}M^{bd}\wedge T^c \wedge I^a - \fr12 H(r)\e_{bcd}M^{ab}\wedge T^c \wedge T^d\bigg), 
    \end{aligned}
\end{equation}
where $M^{ab}$ is the boost/rotation current defined as
\begin{equation}
    M^{ab} = \s^{[a}T^{b]}.
\end{equation}

To get rid of the term with extra factor $H(r)$ in the flow equation one should consider special membrane configurations, as we did in the D3-brane case, that imply $\det T= 0$. This can be interpreted as vanishing of one or two eigenvalues or as the energy-momentum tensor being null. The latter case corresponds to a null wave propagating on a static membrane. One vanishing eigenvalue separates one tensionless direction, that can be interpreted as the M2-brane wrapping a KK circle. In this case we get back to the fundamental string. Another possibility can be that there exists an effectively one-dimensional excitation inside the membrane, that could be boundary of another membrane. However, in this case it is not clear how to disentangle dynamics of the additional membrane both from the membrane in question and from the background. Finally, two vanishing eigenvalues mean that the membrane is effectively point-like and is described in terms of a (set of) D0-brane(s). In this case $\dt_\g S = 0$, that can be understood as absence of flow for deformations of particle like M2-brane configurations in 11 dimensions. In Section \ref{sec:coord} we discuss the absence of flow of the action of the uni-vector deformed massless particle in 11 dimensions, that is the parent system for the D0-brane. However, at this point it is not clear, whether the membrane whose energy-momentum tensor has two vanishing eigenvalues on a tri-vector deformed background is equivalent to this system. 

\section{Flows and coordinate transformations}
\label{sec:coord}

In Section \ref{sec:D0} we derived flow equation for the D0-brane world-volume theory that corresponds to uni-vector deformations, that appears to have the same form as in \cite{Blair:2024aqz}. On the other hand in \cite{Gubarev:2025hvr} it was shown that uni-vector deformations in a $D$-dimensional theory are equivalent to coordinate transformations in its parent $D+1$-dimensional theory. In particular, the deformation considered in Section \ref{sec:D0} is equivalent to the coordinate transformation
\begin{equation}
    t \to t + \a\, \psi,
\end{equation}
where $\psi$ is the KK cycle. The parent 11-dimensional action, that is the action for a massless particle, is apparently invariant under such transformations and therefore its derivative along $\a$ vanishes and no flow equation appears. 

Nevertheless, the 10-dimensional D0-brane action has a non-trivial flow, that appears to be due to the mixing between the KK-circle $\psi$ and a remaining coordinate ($t$ in this case). Indeed, as it was shown explicitly  in \cite{Barakin:2026mxz} to eliminate $\psi$ from the massless particle on a KK-reduced background one solves the KK momentum conservation law $p_\psi = $const. This introduces coupling of the KK vector $A_m$ to the charge $q=p_\psi$. Precisely this conservation law is what changes under the deformation, leading to change in the charge $q \to q + \a E$, where $E$ is energy of the particle, and finally to a flow.

It is possible to develop a similar understanding of flows corresponding to bi-vector deformations. For that one starts with the double sigma-model formulation of the strings' dynamics
\begin{equation}
    S = \fr{1}{4\p \a'} \int d^2 \s \Big[ \dt_1 X^M \h_{MN}\dt_0 X^N - \dt_1 X^M \mH^{MN}\dt_1 X^N \Big],
\end{equation}
where $X^M = (X^m, \tX_m)$. Since one has twice more degrees of freedom, the action must be endowed with a self-duality condition that is
\begin{equation}
    \h_{MN}\dt_0 X^N = \mH_{MN}\dt_1 X^N.
\end{equation}
Solving it explicitly one recovers the standard string action. Now, as it has been shown in \cite{Sakamoto:2017cpu,Catal-Ozer:2019tmm,Gubarev:2025hvr} abelian and almost-abelian bi-vector deformations are equivalent to $\rmO(10,10)$ transformations, that can be absorbed into coordinate transformations
\begin{equation}
    X^M \to O^M{}_N X^N, \quad ||O^M{}_N||\in \rmO(10,10).
\end{equation}
The action is apparently invariant under such transformations and in this form does not develop a flow when the underlying background is deformed.

The difference between the uni-vector case, where the absence of flow is visible already in the 10D theory, and the bi-vector case, where in 10D formulation the flow is non-trivial, is in the self-duality condition. Indeed, to go from the $\rmO(10,10)$ invariant formulation to the standard string action one has to decide which coordinates are not-physical and eliminate the corresponding unphysical coordinates. This step is the same as the standard KK reduction. However, now dynamics of dual coordinates $\tX_m$ is related to that of the standard coordinates $X^m$ by
\begin{equation}
    \dt_0 \tX_m = g_{mn}\dt_1 X^n + B_{mn} \dt_0 X^n.
\end{equation}
This relation changes upon a bi-vector deformation and, after substituting into the action, results in an explicit dependence on the deformation parameter. In other words, when solving the self-duality condition one has to decide which coordinates are understood as non-geometric and will be removed from the action. This is similar to how different branes of string theory can be obtained from a single O(10,10)-covariant action \cite{Blair:2017hhy,Bergshoeff:2019sfy,Musaev:2022yqh}. Deformation, understood as a coordinate transformation in doubled space-time, mixes dual and normal coordinates and thus changes this identification. At the level of the string action this results in a dependence on the deformation parameter expressed as a flow equation. It is tempting to relate the integrability property of the $\mathrm{T}\bar{\mathrm{T}}$ flow and similar deformations to the fact that they are nothing but coordinate transformations in an appropriate formulation. Preservation of classical integrability of the 2D system itself under TsT bi-vector deformations was made explicit in \cite{Orlando:2019rjg}, where constant $\rmO(10,10)$ transformations understood as doubled coordinates transformations were considered. They explicitly show that Lax connection is preserved. It is therefore interesting to extend their analysis to the non-abelian deformations considered here.

\section{Discussion}
\label{sec:conclusions}

In this paper we analysed the impact of poly-vector deformations of the underlying background at world-volume theories of the fundamental string, D0- and D3-branes and the 11-dimensional membrane. We consider deformations generated by bi-vectors, uni-vectors, quadri-vectors and tri-vectors respectively. For the initial backgrounds we take the fundamental string, the D0- and D3-brane, and the M2 backgrounds. In the case of the probe string we also consider the $\AdS_3\times \SS^3 \times \TT^4$ background. For the poly-vectors we take Killing vectors of the AdS part of the near-horizon region of each brane background. These are not isometries of the full background, however, some poly-vectors still generate solutions. To include special conformal transformations we consider the $\AdS_3\times \SS^3 \times \TT^4$ solution, of which they are proper Killing vectors. In all other cases they do not generate solutions.

For all considered setups, both for abelian and non-abelian deformations, we observe the same behaviour of the world-volume action: a deformation of the underlying geometry manifests itself as a flow of the world-volume action of a probe with the deformation parameter. More concretely, for bi-vector deformations of the form $P\wedge A$, where $P$ stands for a momentum generator and $A$ is a placeholder for $D$, $M$ or special conformal generators, the probe action satisfies
\begin{equation}
    \fr{\dt S}{\dt \g_\pm} \sim \int d^2\s j_A\wedge T_\pm.
\end{equation}
Here, $j_A$ is the 2d QFT 1-form current (non-conserved after deformation) corresponding to the generator $A$ and 
\begin{equation}
    \begin{aligned}
        T^+ & = T_{++}dx^+ + T_{+-}dx^-, \\
        T^- & = T_{+-}dx^+ + T_{--}dx^-.
    \end{aligned}
\end{equation}
For the membrane on tri-vector deformed 11D background we find a similar expression involving wedge products of three 1-form current, each corresponding to generators entering the tri-vector. For the D0- and D3-branes we were unable to rewrite the flow in the same form for particular deformations, however, the general flow equation takes the same structural form as that of \cite{Blair:2024aqz}. In general, the flow equation for the D3-brane contains a square-root term in the RHS. The difference is in a non-trivial defining function $H(r)$ and the non-abelian nature of deformations.

In Section \ref{sec:coord} we discuss the results in the prism of the work \cite{Gubarev:2025hvr}, where uni-vector and (almost-abelian) bi-vector deformations were shown to be equivalent to coordinate transformations. The transformation is in the 11D and doubled space-time respectively. Important is, that the parent action, say the massless particle in 11D for the D0 case, stays invariant under such transformations and therefore no flow should appear. A non-trivial flow equation for the world-volume action is the result of a particular KK reduction scheme, that is changed under the coordinate transformation. The same mechanism persists for bi-vectors at the level of the doubled sigma model. There, a bi-vector deformation is likewise trivial before choosing a physical polarization. The non-trivial flow of the ordinary sigma-model appears only after solving the self-duality constraint, i.e. after selecting which ten coordinates out of the doubled set $X^M$ are to be eliminated. It is therefore natural to conjecture that exact solvability of such flows, and as a consequence integrability of the deformation, is a result of its equivalence to coordinate transformation. Hence, one may expect that non-abelian flows analysed here might also be integrable as flow equations. For example, an exact expression for energy spectrum of the probe string should exist.

This stands among interesting directions for further extension of the presented results. Another interesting question is whether the correspondence between the form of a deformation bi-vector and the flow is a general feature. In other words, whether it is true, that for a bi-vector $A_1\wedge A_1$ the flow is controlled by $j_{A_1}\wedge j_{A_2}$. Finally, as we mention at the end of Section \ref{sec:M2},  a more detailed analysis of the M2 flow equation seems to be possible. In particular, it is not completely clear, why the flow disappears for such membrane configurations that the world-volume energy-momentum tensor has two vanishing eigenvalues. Intuitively, these are related to a particle-like description of M2, and its relation to D0-branes is therefore required.

\section*{Acknowledgments}

The work of A.K. has been supported by the Theoretical Physics and Mathematics Advancement Foundation “BASIS” grant 25-2-1-111-1.

\appendix 

\section{Bi-vectors of the string background}
\label{app:class}

We solve the classical Yang--Baxter equation (CYBE), whose solutions are antisymmetric $r$-matrices. Let us enumerate generators of SO(2,2) as follows
\begin{equation}
    \{T_A\} = (P_1,P_2,M,D,K_1,K_2).
\end{equation}

Let us start with classifying solutions to the Yang--Baxter equation with no contribution from the generators of special conformal transformation. Such a bi-vector then takes the form
\begin{equation}
    \b = \r_{12}\,P_1\wedge P_2 + \r_{13}\,P_1\wedge M + \r_{2,3}\,P_2\wedge M + \r_{1,4}\,P_1\wedge D+\r_{2,4}\,P_2\wedge D+\r_{3,4}\,M\wedge D,
\end{equation}
and is defined by the $4\times 4$ antisymmetric matrix $r=||\r_{ab}||$. Whether the bi-vector $\b$ can be represented as $
\b=x\wedge y$, that we refer to as the simple bi-vector, is controlled by the Pfaffian $\mathrm{Pf}(r)$ of the corresponding r-matrix 
\begin{equation}
\mathrm{Pf}(r)=\rho_{12}\rho_{34}-\rho_{13}\rho_{24}+\rho_{14}\rho_{23}.
\end{equation} 
The bi-vector $\b$ is simple if $\Pf(r)=0$, that for the $4\times 4$ r-matrix implies $\mathrm{rank}( r)=2$. If $\Pf(r)\neq 0$, then $\mathrm{rank}(r)=4$, and $\b$ can only be represented as a sum of simple bi-vectors. Let us list all possibilities in this sector of bi-vectors.

For vanishing Pfaffian we have:
\begin{enumerate}
    \item $    r_1 = \Bigl(p_1-\frac{\rho_{2,4}}{\rho_{1,2}}\,d\Bigr)\wedge
    \Bigl(\rho_{1,2}p_2+\rho_{1,4}d\Bigr)$; unimodularity gives  $r=\r_{1,2} p_1\wedge p_2$
    \item  $
r_2 = (d-M)\wedge\Bigl(\rho_{1,3}(p_1-p_2)-\rho_{3,4}d\Bigr)$; unimodular, however  supergravity equations hold if $\rho_{3,4}=0$
 \item 
$
r_3 = (M+d)\wedge\Bigl(\rho_{3,4}d-\rho_{1,3}(p_1+p_2)\Bigr)$, unimodular
\item $
r_4 = \Bigl(p_1-\frac{\rho_{2,3}}{\rho_{1,2}}\,M-\frac{\rho_{1,3}}{\rho_{1,2}}\,d\Bigr)\wedge
\Bigl(\rho_{1,2}p_2+\rho_{1,3}M+\rho_{2,3}d\Bigr)$, unimodular;
\item $
r_5 =  (p_1-p_2)\wedge\Bigl(\rho_{1,2}p_2+\rho_{1,3}M+\rho_{1,4}d\Bigr)$, unimodularity gives $r_2$
\item $
r_6 = (p_1+p_2)\wedge\Bigl(\rho_{1,2}p_2+\rho_{1,3}M+\rho_{1,4}d\Bigr)$, unimodularity gives $r_3$
\end{enumerate}

For non-vanishing Pfaffian we have the following solutions
\begin{enumerate}
\item $
r_7 = \rho_{1,2}\,p_1\wedge p_2+\rho_{1,3}\,(p_1\wedge M-p_2\wedge d)$,  $\Pf(r_7) = \rho_{1,3}^2$, unimodularity requires $\rho_{1,3}=0$;
\item $r_8 = \rho_{1,2}\,p_1\wedge p_2
+\rho_{1,3}\,(p_1\wedge M-p_2\wedge d)
+\rho_{2,3}\,(-p_1\wedge d+p_2\wedge M)$ ,  $\Pf(r_8) = \rho_{1,3}^2-\rho_{2,3}^2$, unimodularity requires
$\rho_{1,3}=0$, $\rho_{1,4}=0$
\end{enumerate}

In terms of realization of the $\frso(2,2)$ generators as isometries of the $\AdS_3$ space-time in Poincar\'e coordinates, the bi-vectors listed above are of order 0, 1 and 2 in powers of coordinates. At order 2 one also has bi-vectors of the form $P\wedge K$. Of these there are only two solutions to CYBE, both simple:
\begin{enumerate}
    \item $r_{9} = \rho_{1,5} (p_0 + p_1)\wedge (k_0 + k_1)$, unimodular; 
    \item $r_{10} = \rho_{1,5} (p_0 - p_1)\wedge (k_0 - k_1)$, unimodular.
\end{enumerate}

At order 3 we have bi-vector of the form $d\wedge k + M\wedge k$. Bi-vectors with r-matrix satisfying $\Pf(r) = 0$ are the following 
\begin{enumerate}
    \item $r_{11} = d\wedge (\rho_{4,5} k_1+\rho_{4,6} k_2)$, unimodularity requires $r_{11}=0$;
    \item $r_{12} = (\rho_{4,5} d + \rho_{3,5} M)\wedge ( k_1-k_2)$, unimodularity requires $\r_{4,5} = \r_{3,5}$;
    \item $r_{13} = (\rho_{4,5} d + \rho_{3,5} M)\wedge ( k_1+k_2)$, unimodularity requires $\r_{4,5} = -\r_{3,5}$.
\end{enumerate}
Bi-vectors with non-vanishing Pfaffian read
\begin{enumerate}
    \item $r_{15} = \rho_{3,5} \left(d\wedge k_2+M\wedge k_1\right)+\rho_{4,5} \left(d\wedge k_1+M\wedge k_2\right) $, $\Pf(r_{15}) = \rho_{4,5}^2-\rho_{3,5}^2$, unimodularity requires $r_{15} = 0$;
    \item $r_{16} = \rho_{3,5} \left(d\wedge k_2+M\wedge k_1\right)$, $\Pf(r_{16}) = -\rho_{3,5}^2$, unimodularity requires $r_{16} = 0$.
\end{enumerate}
At order 4, where the only bi-vector is $k_0\wedge k_1$ one finds no solutions to CYBE.

Certainly, one may consider bi-vector of mixed order, however, general analysis in this case is quite technical and stands beyond the scope of our paper.

\bibliography{bib.bib}
\bibliographystyle{utphys.bst}

\end{document}